# Can violations of Bell's inequalities be considered as the final proof of quantum physics ?

François Hénault

Institut de Planétologie et d'Astrophysique de Grenoble
Université Joseph Fourier, Centre National de la Recherche Scientifique
B.P. 53,  38041 Grenoble – France

**ABSTRACT**

Nowadays, it is commonly admitted that the experimental violation of Bell's inequalities that was successfully demonstrated in the last decades by many experimenters, are indeed the ultimate proof of quantum physics and of its ability to describe the whole microscopic world and beyond. But the historical and scientific story may not be envisioned so clearly: it starts with the original paper of Einstein, Podolsky and Rosen (EPR) aiming at demonstrating that the formalism of quantum theory is incomplete. It then goes through the works of D. Bohm, to finally proceed to the famous John Bell's relationships providing an experimental setup to solve the EPR paradox. In this communication is proposed an alternative reading of this history, showing that modern experiments based on correlations between light polarizations significantly deviate from the original spirit of the EPR paper. It is concluded that current experimental violations of Bell's inequalities cannot be considered as an ultimate proof of the completeness of quantum physics models.

**Keywords:** EPR paradox, Bell's inequalities, photon, polarization, correlation

## 1 INTRODUCTION

The publication of the Einstein, Podolsky and Rosen (EPR) paradox in 1935 [1] initiated a strong debate between A. Einstein, N. Bohr and later physicists about the completeness of quantum physics theory. It is only in 1964 when J. S. Bell proposed a realistic experimental test providing us with a quantitative criterion, being able to reconcile all parties [2]. Since more than thirty years, experimental results from worldwide optical benches aiming at testing Bell's inequalities led to impressive results, tending to demonstrate the validity and completeness of Quantum Mechanics Theory [3-7]. It seems however that this story can be rewritten in a different way along the years. This is the main goal of this paper: in section 2 is first described a heuristic, semi-classical model of modern J. Bell's experiments based on photon polarization correlations. Section 3 is an attempt to revisit the science story, going from the original EPR paradox through the works of D. Bohm and J. Bell, to finally conclude on modern experimental results. Section 4 gives a quick summary of the whole study.

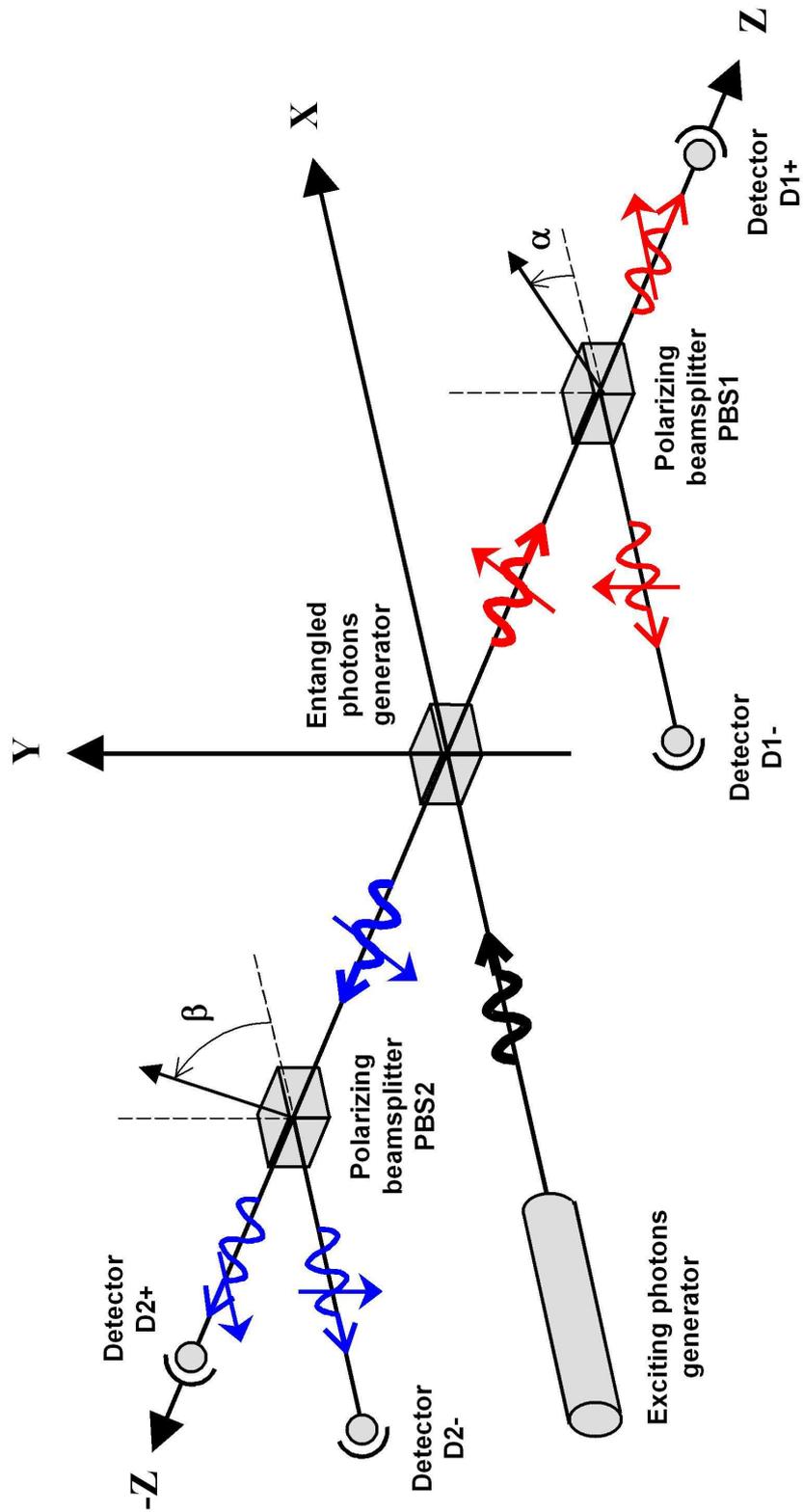

Figure 1: Conceptual view of modern Bell's experiments.

## 2 SEMI-CLASSICAL VIEW OF BELL'S EXPERIMENTS

In this section is presented a semi-classical view of modern J. Bell's experiments (§ 2.1) followed by a discussion about their relevance (§ 2.2).

### 2.1 Heuristic model of contemporary Bell's experiments

In Figure 1 is sketched a conceptual view of modern Bell's experiments based on intricated photons polarization. This is a very general scheme, not taking into account practical implementations such as described in Refs. [3-7] for instance. It basically consists in an exciting photons generator such as a pumping laser and an entangled photon generator based for example on atomic cascades or parametric down-conversion crystals. Then a pair of intricated photons is emitted along opposite directions, say photon n°1 along the +Z axis and photon n°2 along the -Z axis (see the definitions of axes in Figure 1 where photon n°1 goes to the right and photon n°2 to the left). Along each arm are set polarizing beamsplittters (PBS1 and PBS2) and a couple of detectors, respectively denoted [D1+, D1-] and [D2+, D2-]. Depending on the polarization states of the impinging photons and playing with the orientations of PBS1 and PBS2 around the Z-axis (denoted $\alpha$ and $\beta$ in Figure 1 and Figure 2), J. Bell and its followers demonstrated that it was feasible to solve the EPR paradox by computing correlations between counts measured by the four detectors and varying the orientations angles $\alpha$ and $\beta$ of PBS1 and PBS2. Hence it should be possible to conclude about the completeness of Quantum Mechanics theory from real experimental results. But an alternative, heuristic interpretation of these results is as follows.

We assume that right from the generation of one couple of intricated photons, their polarization states are linked, not by "hidden variables" but rather by a "hidden rule" stating that their polarization states must be opposite, according to the universal law of conservation of cinematic momentum[1]. This means for example that an emitted linearly polarized photon having a roll angle $\lambda$ around the +Z-axis will have a counterpart propagating along the -Z-axis under a roll angle $-\lambda$, and that their following optical adventures should strictly comply with the starting rule. This is illustrated in the left panel of Figure 2. Since we are dealing here with autocorrelation relationships, the opposite sign between $\lambda$ and $-\lambda$ can be omitted without loss of generality (i.e. the original parallelism or anti-parallelism between polarized photons 1 and 2 does not matter). Here $\lambda$ may be considered as the hidden variable mentioned in the original Bell's paper [2]. We then assume that the joint detection probabilities between each couple of detectors can be written as:

$$[D1+, D2+] \quad P_{++} = \left| \int_0^{2\pi} \cos(\alpha-\lambda)\cos(\beta-\lambda) \, d\lambda \bigg/ 2\pi \right|^2 = \cos^2(\alpha-\beta)/4$$

$$[D1-, D2-] \quad P_{--} = \left| \int_0^{2\pi} \sin(\alpha-\lambda)\sin(\beta-\lambda) \, d\lambda \bigg/ 2\pi \right|^2 = \cos^2(\alpha-\beta)/4$$

$$[D1+, D2-] \quad P_{+-} = \left| \int_0^{2\pi} \cos(\alpha-\lambda)\sin(\beta-\lambda) \, d\lambda \bigg/ 2\pi \right|^2 = \sin^2(\alpha-\beta)/4$$

$$[D1-, D2+] \quad P_{-+} = \left| \int_0^{2\pi} \sin(\alpha-\lambda)\cos(\beta-\lambda) \, d\lambda \bigg/ 2\pi \right|^2 = \sin^2(\alpha-\beta)/4$$

Eqs. (1)

---

[1] Admitted in both the general relativity and quantum physics theories.

Those expressions are in accordance with Quantum Mechanics Theory [8]. Furthermore, they allow us to compute the correlation coefficient E(α,β) for measurements along α and β, that is [4]:

$$E(\alpha,\beta) = \frac{P_{++} + P_{--} - P_{+-} - P_{-+}}{P_{++} + P_{--} + P_{+-} + P_{-+}} = \cos 2(\alpha - \beta) \qquad \text{Eq. (2)}$$

This again complies with the results of quantum physics. Thus it might be concluded that our semi-classical model also authorizes a violation of Bell's inequalities (see the next subsection for details).

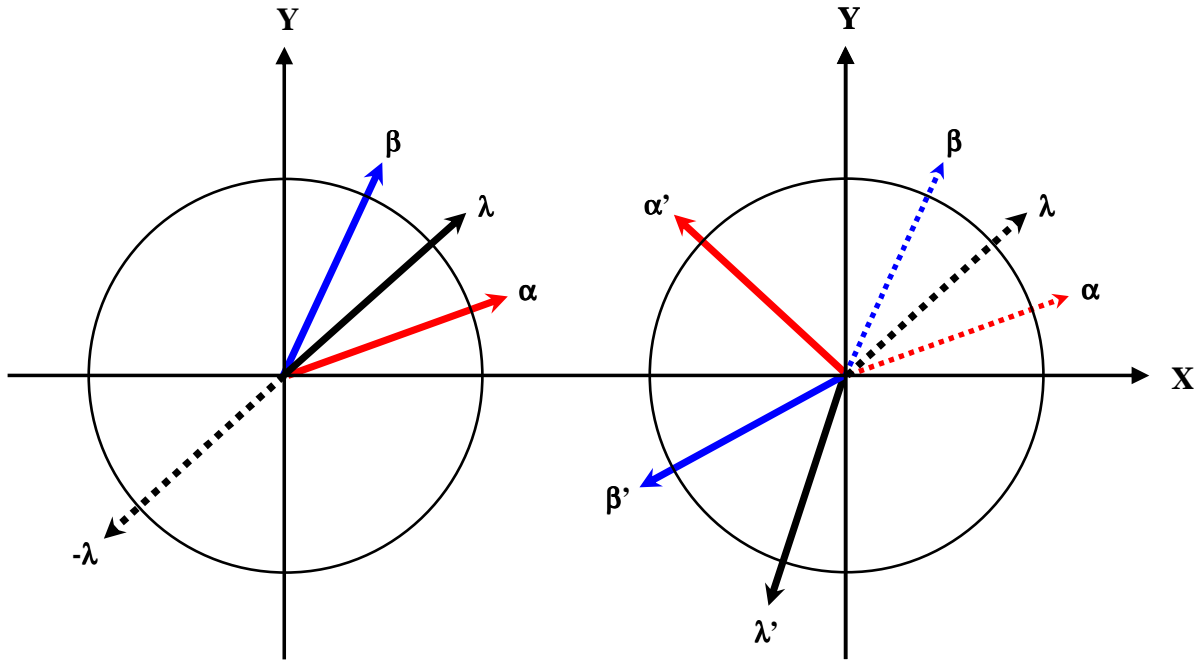

Figure 2: Polarization orientations in modern Bell's experiments.

## 2.2 Are Bell's inequalities applicable?

The mathematical developments in the previous sub-section may at first sight seem to be in contradiction with J. Bell's inequalities, since they also rely on a semi-classical view of particle physics. In fact, the modern expression of Bell's inequalities can be written as follows [4][1]:

$$-2 \leq S(\alpha,\beta,\alpha',\beta') = E(\alpha,\beta) - E(\alpha,\beta') + E(\alpha',\beta) + E(\alpha',\beta') \leq +2, \qquad \text{Eq. (3)}$$

where the couples of angles (α,β) and (α',β') indicate different and *successive* orientations of PBS1 and PBS2 around the Z-axis (see the right panel of Figure 2). But clearly the mathematical relationships Eqs. (1-2), being compliant with Quantum Mechanics, are authorizing a violation of the previous inequality:

$$-2\sqrt{2} \leq S(\alpha,\beta,\alpha',\beta') \leq +2\sqrt{2}, \qquad \text{Eq. (4)}$$

---

[1] It must be noticed that the formalism of Bell's inequalities has been rewritten and expanded many times along the years. A good summary of this story can be found in Ref. [8], Chapter 4.

for certain choices of the angles α, β, α' and β', e.g. α-β = β-α' = α'-β' = 22.5 degs. or α-β = β-α' = α'-β' = 67.5 degs. [3]. Hence the results of modern Bell's experiments could be explained in a purely semi-classical way compatible with quantum physics.

This surprising conclusion may be understood as follows: in the original J. Bell's paper [2] as well as those written by his followers [9-10], the parameter λ is considered as a hidden variable staying constant whatever are the orientations of polarizing beamsplitters PBS1 and PBS2. It leads to computing integral sums of the following type:

$$S(\alpha,\beta,\alpha',\beta') = \int_\lambda \rho(\lambda) \left[E(\alpha,\beta,\lambda) - E(\alpha,\beta',\lambda) + E(\alpha',\beta,\lambda) + E(\alpha',\beta',\lambda)\right] d\lambda, \qquad \text{Eq. (5)}$$

where $\rho(\lambda)$ is the density probability function associated to the hidden variable λ. But according to our semi-classical model described in section 2.1, λ actually is the roll angle of a couple of *simultaneously* emitted photons and should not be considered as a constant parameter for each couple of polarizer orientations (α, β) and (α',β'). Instead we could rewrite Eq. (5) as:

$$S(\alpha,\beta,\alpha',\beta') = \int_\lambda \rho(\lambda) \left[E(\alpha,\beta,\lambda) + E(\alpha',\beta',\lambda)\right] d\lambda + \int_\lambda \int_{\lambda'} \rho'(\lambda,\lambda') \left[-E(\alpha,\beta',\lambda,\lambda') + E(\alpha',\beta,\lambda,\lambda')\right] d\lambda\, d\lambda', \qquad \text{Eq. (6)}$$

where $\rho(\lambda)$ and $\rho'(\lambda,\lambda')$ are different functions because the measurements cannot be *simultaneous*. What would become J. Bell's inequality in that alternative framework remains to be demonstrated. In the mean time, we shall revisit the whole science story, starting from the original EPR paper. This is the purpose of the next section.

## 3 REVISITING THE SCIENCE STORY

Perhaps more interestingly than the semi-classical model of Bell's experiments described in section 2, is a new, alternative reading of the history of the EPR paradox, being based on the original writings of the major implicated authors. This is the main purpose of the present section where some decisive steps are summarized on the flow-chart of Figure 3.

### 3.1 The EPR paradox

Throughout the years, the EPR paradox became the subject of many interpretations, either in textbooks or journal papers. Those include considerations about causality, non-locality, or propagation of information faster than the speed of light (see for example Ref. [8]). Here we just aim at returning to the original text [1], and seeing if it can be understood in simpler ways. The paper starts by near from philosophy considerations: any *complete* and *realistic* physical theory should be able to predict all the results of an experiment undisturbed by the measurement apparatus. From this assumption the authors address the impossibility of *simultaneously* determining the position and momentum of a given particle (that is the Heisenberg's uncertainty principle), and state that:

1) Either the laws of Quantum Mechanics are not complete, i.e. they do not provide us with a full view of physical reality (e.g. *simultaneous* position and momentum of a given particle).

2) Either they are complete. In that case we must accept the uncertainty principle (non commuting parameters cannot be determined *simultaneously*) and therefore sacrifice our usual notions about objective physical reality.

The authors are clearly in favor of the first hypothesis. To demonstrate it, they imagined a *Gedanken* experiment where a couple of identical particles having previously interacted together are propagating in different directions with opposite positions and momentums. Nowadays such particles are said to be "intricated". The authors claim that using an experimental device measuring either the position or momentum of the first particle, it would be possible to characterize the same physical parameters for the second particle with the same accuracy, *simultaneously* and *without disturbing it*, henceforth overcoming the Heisenberg's principle and the whole Copenhagen interpretation of quantum physics. Their conclusion is that the results of this *Gedanken* experiment should contradict hypothesis n°2, hence only hypothesis n°1 remains valid and the theory of Quantum Mechanics is uncompleted.

It finally seems that the original purpose of the authors was to question the validity of the uncertainty principle, which is one of the strongest pillar of quantum physics. Moreover, it must be noticed that the EPR paper never mentions any reference to hidden variables or to the violation of special relativity[1]: those notions emerged later. The very last sentence of the paper simply mentions that the possibility of developing a more complete theory than Quantum Mechanics may exist. Remarkably, a similar idea was evoked by M. Born in the end of a decisive paper about quantum theory [111].

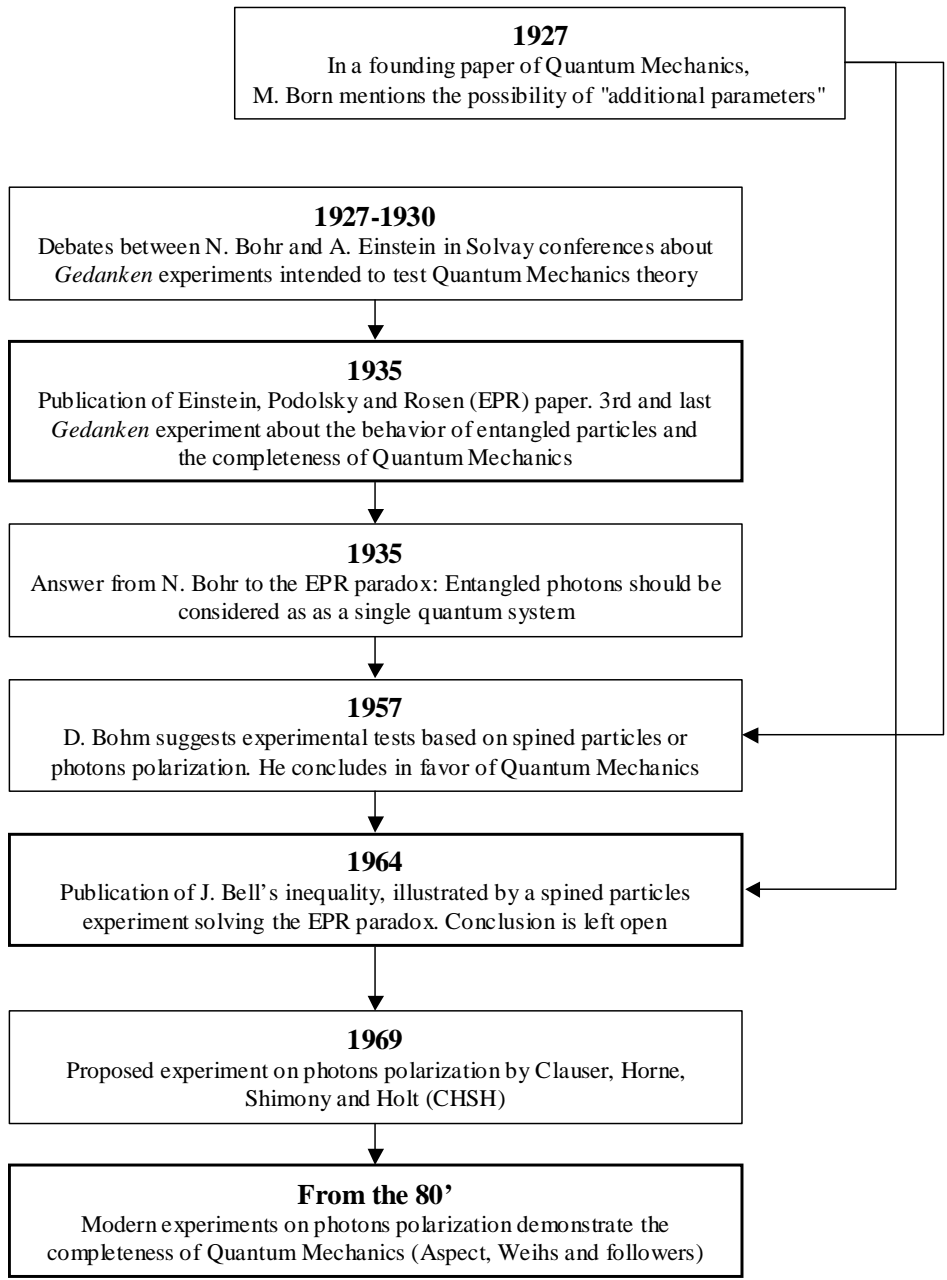

Figure 3: Tentative history of the EPR paradox.

---

[1] The paper simply states that both systems are "independent".

### 3.2 Previous debates between A. Einstein and N. Bohr

Indeed, the EPR paper was not the first attempt from A. Einstein to overcome Heisenberg's uncertainty principle. This story is well-known (see for example Refs. [12-13]) and is only shortly summarized herein.

- The double slit experiment (1927 Solvay conference, left side of Figure 4): this is a variant of Young's experience where the slits are hanged by springs. Classically, the apparatus should allow the *simultaneous* and *accurate* determinations of the position x of an impinging photon particle, and of its momentum p *via* the recoil of the slits. Therefore Heisenberg's uncertainty principle $\Delta x\, \Delta p \geq h/2\pi$ would not be respected[1]. N. Bohr objected that the measurement accuracy of slits position or momentum would not allow such violation, and that if they did, Young's fringes would be destroyed according to Bohr's own principle of complementarity [12].

- The "photon box" experiment (1930 Solvay conference, right side of Figure 4): here A. Einstein contests the second aspect of the uncertainty principle stating that $\Delta E\, \Delta t \geq h/2\pi$, where E and t respectively stand for energy and time. The basic idea was to measure precisely the emission times of photons emitted from a shuttered box and to crosscheck them with mass or energy losses from the box. But here again N. Bohr refuted the argument by invoking Einstein's own theory of relativity about the accuracy of absolute time measurements.

Hence the EPR paradox should be considered as the third and last attack from A. Einstein against Heisenberg's uncertainty principle. The major differences with respect to both previous *Gedanken* experiments is that it was for the first time written in a regular journal paper, and strictly following the mathematical formalism of Quantum Mechanics wave functions. In that prospect, it is very interesting to read again the response of N. Bohr to the EPR paper [14]. He answers that the two intricated particles cannot be considered as independent, since they together form a single system, to which the uncertainty principle is still applicable. It may be noticed that this demonstration is only developed as a footnote in the original text[2], while the main *corpus* of the paper is focused on discussions about the 1927 and 1930 *Gedanken* experiments, and proposes a variant of the 1927 double slit experience to test the EPR paradox. Since his arguments are almost based on photon particles, it may be conjectured that the paper had later influence on modern Bell's experiments.

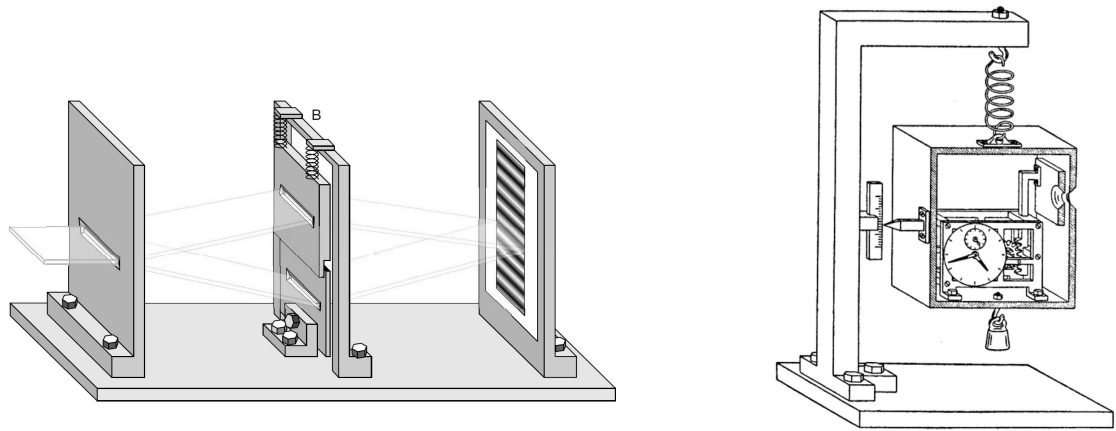

Figure 4: Einstein's *Gedanken* experiments about the recoiling double slit (left) and photon box (right).

### 3.3 D. Bohm's interpretation

Things could have remained unchanged for a long time, and finally being forgotten. But in 1957 Bohm and Aharonov gave a new interpretation of the EPR paradox [16] based on contemporary quantum formalism. They proposed to replace the original notions of position and momentum of particles with the properties of spinned atoms (denoted A and B) generated by a single molecule. They claimed that no *instantaneous* correlation between the spins of A and B atoms can exist because this would violate the laws of special relativity, hence no "hidden interaction" is possible. They inferred that there should exist a "deeper, sub-quantum mechanical level" inspired from De Broglie's "pilot wave" model. In order to solve the EPR paradox, they described a practically simplified setup measuring correlations between the polarization states of intricated photons produced by positron-electron annihilation. They finally provided some quantitative criteria in favor of quantum physics, but at that time their proposed experiments remained too difficult for practical implementation.

---
[1] $h$ being the Planck's constant.
[2] One year later W. H. Furry published more developed arguments in favor of the completeness of quantum physics [15].

Remarkably, it must be noticed that in two earlier papers [17-18], D. Bohm was the first author to explicitly mention the possibility of "hidden variables" completing the formalism of quantum theory. But he apparently abandoned this idea into his 1957 paper [16] and later works. Nevertheless, these words are nowadays universally recognized and repeated in most of publications dealing with the EPR paradox. It finally looks like if D. Bohm was the actual inventor of modern J. Bell's experiments, rather than EPR themselves.

### 3.4 Testing J. Bell's inequalities

Like the EPR paper itself, J. Bell's inequalities were the scope of many papers and textbooks (e.g. Ref. [8]). We also gave our own interpretation of the inequality in sub-section 2.2, so here are just provided a few additional comments about the original text [2].

1) First of all, it must be noticed that Ref. [2] does not state that the *Gedanken* experiment should necessarily be performed by using properties of photons polarization. Instead, the author mentions Stern-Gerlach magnets detectors enabling to discriminate the spin states of other types of particles.

2) In its conclusion, J. Bell also insists on the necessity to perform "on-flight" measurements, which may be understood as another view of J. Wheeler's delayed choice idea. This was later tested on photons polarization by A. Aspect and his collaborators [5].

3) Finally, it seems that J. Bell was the first to employ again the term "hidden variable" into a scientific publication after the original Bohm's papers of 1952 [17-18], at least from my knowledge.

Following Bell and Bohm's ideas, Clauser, Horne, Shimony and Holt (CHSH) rewrote the original inequalities into their modern form [9-10]: for that they introduced the concept of using four different polarizer orientations $(\alpha, \beta, \alpha', \beta')$[1], and conceived the first experimental testing of the EPR paradox. The rest of the story is well-know: A. Aspect *et al* clearly demonstrated the violation of J. Bell's inequality even when the photons polarization state is modified *in-flight* [3-5]. This is generally considered as a decisive proof in favor of the completeness of Quantum Mechanics theory. Finally, further experiments involving photon injection in optical fibers of several kilometers long [7] conducted to hamper the classical principles of *locality* and *causality*. But still stand two puzzling questions: do experiments based on photon polarizations effectively reflect the original spirit of the EPR paper ? And how do they demonstrate the Heisenberg's uncertainty principle ? Answering to both questions is far to be obvious. The points discussed in the final section.

## 4  CONCLUSION

In this paper was presented a semi-classical interpretation of modern J. Bell's experiments, based on a heuristic model of intricated photons polarization that complies with quantum physics theory. We also examined the science history of the EPR paradox: it looks clear that from year 1927, A. Einstein was challenging the probabilistic model of Quantum Mechanics through its major emblematic symbol, i.e. Heisenberg's uncertainty principle. For that purpose he imagined three different *Gedanken* experiments, the third and last one of them being known today as the "EPR paradox". It must be noticed that none of the three original *Gedanken* experiments proposed by A. Einstein (i.e. recoiling double-slit, photon box, and intricated particles) have ever been realized experimentally under their original form, and still present important practical difficulties. Finally, two important remarks shall be made to conclude the present communication:

---

[1] Instead of three, noted $(\alpha, \beta, \gamma)$ in the original Bell's paper.

1) From an epistemology point of view, it is striking to see how an original communication discussing the achievement of *simultaneous* position and momentum measurements of any type of particle could result in modern experiments based on photon polarizations. Here it seems that the spirit of the original EPR paper was lost throughout the years, thus it is likely that a proof of the completeness of quantum physics theory still remains to be demonstrated.

2) The present paper should not be understood as criticism against modern Bell's experiments, because they conducted to decisive technological advances into the field of single-photon sources. Moreover, it does not contradict any principle of quantum cryptography or future computing capacities.